\documentclass[twocolumn,amsmath,amssymb,superscriptaddress]{revtex4-1}

\usepackage{graphicx}

\usepackage{amstext,amsmath,amssymb,amsfonts}
\usepackage{bm}

\usepackage{txfonts}
\usepackage{mathtools}
\DeclareMathAlphabet{\mathpzc}{OT1}{pzc}{m}{it}
\DeclareFontFamily{OT1}{pzc}{}
\DeclareFontShape{OT1}{pzc}{m}{it}{<-> s * [1.100] pzcmi7t}{}
\DeclareMathAlphabet{\mathpzc}{OT1}{pzc}{m}{it}

\usepackage[T1]{fontenc}
\usepackage[utf8]{inputenc}

\usepackage{hyperref}

\usepackage{color}
\definecolor{lightblue}{rgb}{0.2,0.2,0.7}
\definecolor{darkblue}{rgb}{0,0.25,0.5}
\definecolor{redbrown}{rgb}{0.875,0.25,0.125}
\definecolor{darkgreen}{rgb}{0,0.5,0}

\newcommand{\bra}[1]{\ensuremath{\langle #1 \vert}}
\newcommand{\ket}[1]{\ensuremath{\vert #1  \rangle}}
\newcommand{\braket}[2]{\ensuremath{\langle  #1 \vert #2  \rangle}}
\renewcommand{\b}[1]{\ensuremath{\mathbf{#1}}}
\renewcommand{\H}{\ensuremath{\text{H}}}

\renewcommand{\l}{\ensuremath{\lambda}}

\newcommand{\sr}{\ensuremath{\text{sr}}}
\newcommand{\ee}{\ensuremath{\text{ee}}}

\newcommand{\HF}{\ensuremath{\text{HF}}}

\newcommand{\CBS}{\ensuremath{\text{CBS}}}

\renewcommand{\d}{\ensuremath{\text{d}}}

\newcommand{\x}{\ensuremath{\text{x}}}
\newcommand{\xc}{\ensuremath{\text{xc}}}
\renewcommand{\c}{\ensuremath{\text{c}}}
\newcommand{\Hxc}{\ensuremath{\text{Hxc}}}
\newcommand{\Hx}{\ensuremath{\text{Hx}}}

\newcommand{\B}{\ensuremath{{\cal B}}}
\newcommand{\FCI}{\ensuremath{\text{FCI}}}

\begin{document}

\title{Double-hybrid density-functional theory with density-based basis-set correction}
\author{Aurore Zna{\"i}da}
\author{Julien Toulouse}
\email{julien.toulouse@sorbonne-universite.fr}
\affiliation{Laboratoire de Chimie Th\'eorique, Sorbonne Universit\'e and CNRS, F-75005 Paris, France}

\date{June 20, 2025}

\begin{abstract}
We develop the theory justifying the application of the density-based basis-set correction (DBBSC) method to double-hybrid approximations in order to accelerate their basis convergence. We show that, for the one-parameter double hybrids based on the adiabatic connection, the exact dependence of the basis-set correction functional on the coupling-constant parameter $\l$ involves a uniform coordinate scaling by a factor $1/\l$ of the density and of the basis functions. Neglecting this uniform coordinate scaling corresponds essentially to the recent work of Mester and K\'allay [J. Phys. Chem. Lett. \textbf{16}, 2136 (2025)] on the application of the DBBSC method to double-hybrid approximations. Test calculations on molecular atomization energies and reaction barrier heights confirm that the DBBSC method efficiently accelerates the basis convergence of double-hybrid approximations, and also show that neglecting the uniform coordinate scaling is a reasonable approximation.
\end{abstract}

\maketitle

\section{Introduction}

In density-functional theory (DFT) of molecular electronic systems, an increasingly popular approach is given by double-hybrid approximations~\cite{Gri-JCP-06}, which consist in combining fractions of Hartree-Fock (HF) exchange and second-order M{\o}ller-Plesset (MP2) correlation energies with exchange and correlation density functionals. Double-hybrid approximations are not merely pragmatic approaches but can be thought of as well-defined approximations of an exact-in-principle multideterminant extension of the Kohn-Sham (KS) scheme using the adiabatic-connection formalism~\cite{ShaTouSav-JCP-11,Fro-JCP-11}. Nowadays, double-hybrid approximations are generally recommended for molecular quantum-chemistry calculations~\cite{MarSan-IJC-20,SanBrePerCioAda-ES-22,BurMewHanGri-AC-22}, at least for systems without significant strong (or static) electron correlation effects.

However, one important limitation of double-hybrid approximations is that, due to the MP2 correlation term, they have a slower convergence with respect to the size of the one-electron basis set in comparison with other DFT approximations, thus requiring relatively large basis sets, possibly with basis-set extrapolations~\cite{ChuChe-JCC-11,Kra-JCTC-20}, with the consequence of increased computational costs. To overcome this problem, it has been proposed to use the explicitly-correlated F12 approach in double-hybrid approximations~\cite{KarMar-JCP-11,MehMar-JCTC-22,MehMar-JPCL-22} or to use small basis sets specially tailored for double-hybrid approximations~\cite{GarBreCamCioAda-JCTC-19,LiBreSanAda-RSCA-21}.

Very recently, Mester and K\'allay~\cite{MesKal-JPCL-25} proposed to accelerate the basis-set convergence of double-hybrid approximations by using the density-based basis-set correction (DBBSC) approach~\cite{GinPraFerAssSavTou-JCP-18,LooPraSceTouGin-JPCL-19}. The DBBSC approach consists in adding to the energy a density functional estimating the basis-set incompleteness error for any given basis set. This DBBSC approach was shown to effectively accelerate the basis-set convergence of pure wave-function calculations of energies and properties for a variety of atomic and molecular systems~\cite{GinPraFerAssSavTou-JCP-18,LooPraSceTouGin-JPCL-19,GinSceTouLoo-JCP-19,LooPraSceGinTou-JCTC-20,GinSceLooTou-JCP-20,GinTraPraTou-JCP-21,YaoGinLiTouUmr-JCP-20,YaoGinAndTouUmr-JCP-21,TraTouGin-JCP-22,MesKal-JCTC-23,HesGinReiKnoWerTou-JCC-24,TraTouGin-FD-24,MesNagKal-JCTC-24,MesKal-CPL-25}. Mester and K\'allay showed that the DBBSC approach can be used for double-hybrid approximations, enabling to obtain near complete-basis-set (CBS) results using computationally affordable basis sets.

In the present work, we closely examine the theory behind the application of the DBBSC approach to double-hybrid approximations. We show how to rigorously formulate the basis-set correction within the one-parameter double-hybrid DFT framework of Ref.~\onlinecite{ShaTouSav-JCP-11}. Using uniform coordinate scaling, we find the exact dependence of the basis-set correction density functional on the parameter $\l$ that fixes the amount of electron-electron interaction treated in the wave-function part of the calculation. This exact $\l$-dependence involves a uniform coordinate scaling by a factor $1/\l$ of the density and of the basis functions inside the basis-set correction density functional. In the work of Mester and K\'allay~\cite{MesKal-JPCL-25}, this uniform coordinate scaling is not applied. In the present paper, using small representative test sets of molecular atomization energies and reaction barrier heights, we study the performance of the DBBSC approach to accelerate the basis convergence of some one-parameter double-hybrid approximations, and we check the effect of the uniform coordinate scaling on the basis-set correction functional. We also check the effect of adding a one-electron basis-set correction.

\section{Theory}

\subsection{Density-based basis-set correction}

Let us consider a spatial-orbital basis set $\B=\{\varphi_p \}_{p=1}^M \subset L^2 (\mathbb{R}^3, \mathbb{C})$ and the $N$-electron Hilbert space ${\cal H}^{\B}$ generated by this basis set. The $N$-electron full-configuration-interaction (FCI) ground-state energy is
\begin{eqnarray}
E_\FCI^{\B} = \min_{\Psi \in {\cal W}^\B} \bra{\Psi} \hat{T} + \hat{V}_\text{ne} + \hat{W}_\text{ee} \ket{\Psi},
\end{eqnarray}
where ${\cal W}^\B = \{ \Psi \in {\cal H}^{\B} \; | \; \braket{\Psi}{\Psi}=1 \}$ is the set of normalized $N$-electron wave functions, $\hat{T}$ is the kinetic-energy operator, $\hat{V}_\text{ne} = \int_{\mathbb{R}^3} v_\text{ne}(\b{r}) \hat{n}(\b{r}) \d\b{r}$ is the electron-nuclei interaction operator, expressed with the electron-nuclei potential $v_\text{ne}$ and the density operator $\hat{n}(\b{r})$, and $\hat{W}_\text{ee}$ is the Coulomb two-electron interaction operator. As well known, $E_\FCI^{\B}$ has a slow convergence to the exact ground-state energy $E_0$ as the size of the basis set $\B$ is increased toward a complete basis of the infinite-dimensional Hilbert space $L^2 (\mathbb{R}^3, \mathbb{C})$ (see, e.g., Refs.~\cite{HelKloKocNog-JCP-97,HalHelJorKloKocOlsWil-CPL-98}).

To bypass this slow basis-set convergence, it was proposed in the density-based basis-set correction (DBBSC) scheme~\cite{GinPraFerAssSavTou-JCP-18,LooPraSceTouGin-JPCL-19,GinSceLooTou-JCP-20} to estimate the ground-state energy by
\begin{eqnarray}
E_0^{\B} = \min_{n \in {\cal D}^\B} \left( F[n] + \int_{\mathbb{R}^3} v_\text{ne}(\b{r}) n(\b{r}) \d \b{r} \right),
\label{E0B}
\end{eqnarray}
where ${\cal D}^\B = \{ n \;|\; \exists\; \Psi \in {\cal W}^\B, \; n_\Psi = n\}$ is the set of one-electron densities $n$ coming from a wave function $\Psi \in {\cal W}^\B$ and $F$ is the standard Levy-Lieb universal density functional~\cite{Lev-PNAS-79,Lie-IJQC-83} defined as
\begin{eqnarray}
\forall n \in {\cal D} ^\CBS, \; F[n] = \min_{\Psi \in {\cal W}_n^\CBS} \bra{\Psi} \hat{T} + \hat{W}_\text{ee}  \ket{\Psi},
\end{eqnarray}
where ${\cal W}_n^\CBS = \{ \Psi \in {\cal W}^\CBS\; |\; n_\Psi = n\}$ is the set of wave functions from the CBS-limit wave-function space ${\cal W}^\CBS$ giving the density $n$. The Levy-Lieb universal density functional is defined on the set of $N$-representable densities (in the CBS limit) ${\cal D}^\CBS = \{ n \;|\; \exists\; \Psi \in {\cal W}^\CBS, \; n_\Psi = n\}$. Roughly speaking, as the size of the basis set $\B$ is increased toward the CBS limit, the density has a faster convergence than the many-body wave function, and thus the energy $E_0^{\B}$ converges faster to the exact ground-state energy $E_0$ than the FCI ground-state energy $E_\FCI^{\B}$ does. 

To obtain a more convenient expression for $E_0^{\B}$, the density functional $F$ is decomposed as~\cite{GinPraFerAssSavTou-JCP-18}
\begin{eqnarray}
\forall n \in {\cal D}^\B, \; F[n] = F^{\B}[n] + \bar{E}^{\B}[n],
\label{FdecompB}
\end{eqnarray}
where $F^{\B}$ is the Levy-Lieb density functional where the wave function is restricted to the basis-set wave-function space ${\cal W}^\B$
\begin{eqnarray}
\forall n \in {\cal D}^\B, \; F^{\B}[n] = \min_{\Psi \in {\cal W}_{n}^\B} \bra{\Psi} \hat{T} + \hat{W}_\ee \ket{\Psi},
\label{FB}
\end{eqnarray}
and $\bar{E}^{\B}$ is a basis-set correction density functional defined to make Eq.~(\ref{FdecompB}) exact. Inserting this decomposition in Eq.~(\ref{E0B}), we arrive at the following expression for $E_0^\B$
\begin{eqnarray}
E_0^{\B} = \min_{\Psi \in {\cal W}^\B} \left( \bra{\Psi} \hat{T} + \hat{V}_\text{ne} + \hat{W}_\ee \ket{\Psi} + \bar{E}^{\B}[n_\Psi] \right),
\label{E0BminPsi}
\end{eqnarray}
which defines a FCI ground-state energy calculation in the basis set $\B$ with a self-consistent basis-set correction $\bar{E}^{\B}$~\cite{GinTraPraTou-JCP-21}.

\subsection{Double-hybrid density-functional theory with basis-set correction}

We now show how set up a rigorous double-hybrid DFT within the DBBSC framework. First, we decompose the density functional $F$ as~\cite{ShaTouSav-JCP-11}
\begin{eqnarray}
\forall n \in {\cal D} ^\CBS, \; F[n] = F^\l[n] + \bar{E}_\Hxc^\l[n],
\label{Fmultidethybrid}
\end{eqnarray}
where $F^\l$ is the Levy-Lieb universal density functional along the adiabatic connection~\cite{GunLun-PRB-76,LanPer-SSC-75,LanPer-PRB-77,Har-PRA-84} with a coupling constant $\l \in [0,1]$
\begin{eqnarray}
F^\l[n] = \min_{\Psi \in {\cal W}_{n}^\CBS} \bra{\Psi} \hat{T} + \l \hat{W}_\ee \ket{\Psi},
\end{eqnarray}
and $\bar{E}_\Hxc^\l$ is a complementary density functional defined to make Eq.~(\ref{Fmultidethybrid}) exact. This complementary density functional can be decomposed as
\begin{eqnarray}
\bar{E}_\Hxc^\l[n] = (1-\l) E_\H[n] + (1-\l) E_\x[n] + \bar{E}_\c^\l[n],
\end{eqnarray}
where $E_\H$ and $E_\x$ are the standard Hartree and exchange functionals of KS DFT, and the last correlation contribution can be written as, for $\lambda \neq 0$,~\cite{LevPer-PRA-85,LevYanPar-JCP-85,Lev-PRA-91,LevPer-PRB-93}
\begin{eqnarray}
\bar{E}_\c^\l[n] = E_\c[n] - \l^2 E_\c[n_{1/\l}], 
\label{barEcl}
\end{eqnarray}
where $E_\c$ is the standard correlation functional of KS DFT and $n_{1/\l}(\b{r})= (1/\l)^3 n(\b{r}/\l)$ is the scaled density. Second, we decompose the density functional $F^\l$ as
\begin{eqnarray}
\forall n \in {\cal D}^\B, \; F^\l[n] = F^{\l,\B}[n] + \bar{E}^{\l,\B}[n],
\label{Fldecomp}
\end{eqnarray}
where $F^{\l,\B}$ is the $\l$-dependent Levy-Lieb density functional where the wave function is restricted to the basis-set wave-function space ${\cal W}^\B$
\begin{eqnarray}
\forall n \in {\cal D}^\B, \; F^{\l,\B}[n] = \min_{\Psi \in {\cal W}_{n}^\B} \bra{\Psi} \hat{T} + \l \hat{W}_\ee \ket{\Psi},
\label{FlB}
\end{eqnarray}
and $\bar{E}^{\l,\B}$ is a basis-set correction density functional defined to make Eq.~(\ref{Fldecomp}) exact. Inserting these decompositions in Eq.~(\ref{E0B}) and using the explicit form of $F^{\l,\B}$ in Eq.~(\ref{FlB}), we arrive at the following expression for $E_0^\B$
\begin{eqnarray}
E_0^{\B} = \min_{\Psi \in {\cal W}^\B} \left( \bra{\Psi} \hat{T} + \hat{V}_\text{ne} + \l \hat{W}_\ee \ket{\Psi} + \bar{E}_\Hxc^\l[n_\Psi] + \bar{E}^{\l,\B}[n_\Psi] \right).
\nonumber\\
\label{E0BminPsil}
\end{eqnarray}
Note that, since we have not yet introduced any approximation, Eq.~(\ref{E0BminPsil}) gives the same $E_0^\B$ as in Eq.~(\ref{E0BminPsi}), independently of $\l$.
In the CBS limit, the basis-set correction density functional $\bar{E}^{\l,\B}$ vanishes, and we recover the multideterminant extension of KS DFT of Refs.~\onlinecite{ShaTouSav-JCP-11,ShaSavJenTou-JCP-12} on which a rigorous double-hybrid DFT can be formulated. For a finite basis set $\B$ and coupling constant $\l=1$, the complementary Hartree-exchange-correlation density functional $\bar{E}_\Hxc^\l$ vanishes, and we recover the (self-consistent) DBBSC scheme in Eq.~(\ref{E0BminPsi}).

Following the same ideas as in Refs.~\onlinecite{AngGerSavTou-PRA-05,ShaTouSav-JCP-11}, a double-hybrid DFT with basis-set correction can be developed as a particular approximation to Eq.~(\ref{E0BminPsil}). First, a density-scaled one-parameter hybrid (DS1H) approximation, with basis-set correction, is defined by restricting the minimization in Eq.~(\ref{E0BminPsil}) to single-determinant wave functions,
\begin{eqnarray}
E^{\text{DS1H},\l,\B}_0 &=& \min_{\Phi \in {\cal S}^\B} \Bigl(\bra{\Phi}\hat{T}+\hat{V}_{\text{ne}}+\l\hat{W}_{\ee} \ket{\Phi}+\bar{E}_{\Hxc}^{\l}[n_{\Phi}] 
\nonumber\\
&&\;\;\;\;\;\;\;\;   + \bar{E}^{\l,\B}[n_\Phi]\Bigl), 
\end{eqnarray} 
where ${\cal S}^\B$ is the set of single-determinant wave functions for the basis set $\B$. A minimizing single-determinant wave function $\Phi^{\l,\B}$ must satisfy the self-consistent Schrödinger equation
\begin{eqnarray}
\hat{P}^\B \Bigl( \hat{T}+\hat{V}_{\text{ne}}+\l \hat{V}_{\Hx}^{\HF}[\Phi^{\l,\B}] + \hat{\bar{V}}_{\Hxc}^{\l}[n_{\Phi^{\l,\B}}] 
\nonumber\\
\;\;\;\;\;\;+ \hat{\bar{V}}^{\l,\B}[n_{\Phi^{\l,\B}}] \Bigl) \ket{\Phi^{\l,\B}} = \hat{P}^\B {\cal E}_0^{\l,\B} \ket{\Phi^{\l,\B}},
\label{DS1Heigenval}
\end{eqnarray} 
where $\hat{P}^\B$ is the projector on ${\cal H}^\B$, $\hat{V}_{\Hx}^{\HF}$ is the nonlocal HF potential operator, $\hat{\bar{V}}_{\Hxc}^{\l}$ is the local Hartree-exchange-correlation potential operator generated by the energy functional $\bar{E}_{\Hxc}^{\l}$, $\hat{\bar{V}}^{\l,\B}$ is the local basis-set correction potential operator generated by the energy functional $\bar{E}^{\l,\B}$, and ${\cal E}_0^{\l,\B}$ is an energy eigenvalue. The DS1H ground-state energy can be finally written as
\begin{eqnarray}
E^{\text{DS1H},\l,\B}_0 = \bra{\Phi^{\l,\B}} \hat{T}+\hat{V}_{\text{ne}} \ket{\Phi^{\l,\B}} + E_{\H}[n_{\Phi^{\l,\B}}] + \l E_\x^{\HF}[\Phi^{\l,\B}]
\nonumber\\
+ (1-\l) E_\x[n_{\Phi^{\l,\B}}] + \bar{E}^\l_\c[n_{\Phi^{\l,\B}}] + \bar{E}^{\l,\B}[n_{\Phi^{\l,\B}}], \;\;\;
\label{DS1H}
\end{eqnarray}
where the full Hartree energy $E_{\H}[n]$ has been recomposed. Second, a nonlinear Rayleigh-Schr\"odinger perturbation theory~\cite{AngGerSavTou-PRA-05,FroJen-PRA-08,Ang-PRA-08} starting from this DS1H reference is defined with the following energy expression with the perturbation parameter $\alpha \in [0,1]$,
\begin{eqnarray}
E^{\l,\B,\alpha}_0 &=& \min_{\Psi \in {\cal W}^\B}\Bigl(\bra{\Psi}\hat{T}+\hat{V}_{\text{ne}}+\l \hat{V}_{\Hx}^\HF[\Phi^{\l,\B}]  + \alpha \l \hat{W} \ket{\Psi} 
\nonumber\\
&&\;\;\;\;\;\;\;\;  +\bar{E}_{\Hxc}^{\l}[n_{\Psi}]  + \bar{E}^{\l,\B}[n_\Psi]\Bigl), 
\label{ElBa}
\end{eqnarray}
where $\l\hat{W}=\l \left( \hat{W}_{\ee} - \hat{V}_{\Hx}^\HF[\Phi^{\l,\B}] \right)$ is the scaled M{\o}ller--Plesset perturbation operator. For $\alpha=0$, the stationary equation associated with Eq.~(\ref{ElBa}) reduces to the DS1H eigenvalue equation [Eq.~(\ref{DS1Heigenval})]. For $\alpha=1$, Eq.~(\ref{ElBa}) reduces to Eq.~(\ref{E0BminPsil}), so $E^{\l,\B,\alpha=1}_0 = E_0^{\B}$, independently of $\l$. The sum of the zeroth-order energy and first-order energy correction gives simply the DS1H energy, $E^{\text{DS1H},\l,\B}_0=E^{\l,\B,(0)}_0+E^{\l,\B,(1)}_0$. Thanks to the existence of a Brillouin theorem just like in standard M{\o}ller--Plesset perturbation theory (see Refs.~\cite{AngGerSavTou-PRA-05,FroJen-PRA-08,Ang-PRA-08}), only double excitations contribute to the first-order wave-function correction $\Psi^{\l,\B,(1)}$ and the second-order energy correction has a standard MP2 form
\begin{eqnarray}
E^{\l,\B,(2)}_0 = \l^2 \bra{\Phi^{\l,\B}} \hat{W} \ket{\Psi^{\l,\B,(1)}}= \l^2 E_\c^{\text{MP2},\l,\B},
\label{}
\end{eqnarray}
where $E_\c^{\text{MP2},\l,\B}$ is evaluated with DS1H orbitals and associated orbital eigenvalues (which implicitly depend on $\l$). In total, this defines a density-scaled one-parameter double-hybrid (DS1DH) approximation, with basis-set correction, in which the exchange-correlation energy contribution is
\begin{eqnarray}
E^{\text{DS1DH},\l,\B}_{\xc} &=& \l E_\x^{\HF}[\Phi^{\l,\B}] + (1-\l) E_\x[n_{\Phi^{\l,\B}}] 
\nonumber\\
&& + \bar{E}^\l_\c[n_{\Phi^{\l,\B}}] + \l^2 E_\c^{\text{MP2},\l,\B} + \bar{E}^{\l,\B}[n_{\Phi^{\l,\B}}].
\label{ExcDS1DH}
\end{eqnarray} 
If we neglect the density scaling in the correlation functional, i.e. $E_\c[n_{1/\l}]\approx E_\c[n]$ in Eq.~(\ref{barEcl}), we can also define a one-parameter double-hybrid (1DH) approximation, with basis-set correction, 
\begin{eqnarray}
E^{\text{1DH},\l,\B}_{\xc} = \l E_\x^{\HF}[\Phi^{\l,\B}] + (1-\l) E_\x[n_{\Phi^{\l,\B}}] \;\;\;\;\;\;\;\;\;
\nonumber\\
 (1-\l^2)E_\c[n_{\Phi^{\l,\B}}] + \l^2 E_\c^{\text{MP2},\l,\B} + \bar{E}^{\l,\B}[n_{\Phi^{\l,\B}}],
\label{Exc1DH}
\end{eqnarray}
in which the fraction of HF exchange is $\l$ and the fraction of MP2 correlation is $\l^2$. These are extensions of the DS1DH and 1DH schemes of Ref.~\onlinecite{ShaTouSav-JCP-11} to a finite basis set $\B$ and with the addition of the basis-set correction density functional $\bar{E}^{\l,\B}$.

\subsection{The basis-set correction density functional $\bar{E}^{\l,\B}$}

\subsubsection{Exact scaling relation for $\bar{E}^{\l,\B}$}

We first want to determine how the basis-set correction density functional $\bar{E}^{\l,\B}$ depends on $\l$. This can be found by generalizing the uniform coordinate scaling relation known for the functional $F^\l$~\cite{LevPer-PRA-85,Lev-PRA-91,Lev-INC-95}. 

For any wave function $\Psi \in {\cal W}_n^\B$, we introduce the scaled wave function $\Psi_\gamma$ defined as (omitting the untouched spin coordinates)
\begin{eqnarray}
\Psi_\gamma(\b{r}_1,...,\b{r}_N)  = \gamma^{3N/2} \Psi(\gamma\b{r}_1,...,\gamma\b{r}_N),
\label{Psig}
\end{eqnarray}
where $\gamma>0$ is a scaling parameter. This scaled wave function $\Psi_\gamma$ yields the scaled density $n_\gamma(\b{r}) = \gamma^3 n(\gamma \b{r})$ and belongs to the Hilbert space generated by  the scaled basis set $\B_\gamma = \{\varphi_{p,\gamma} \}_{p=1}^M $ made of the scaled spatial orbitals $\varphi_{p,\gamma}(\b{r}) = \gamma^{3/2} \varphi_{p}(\gamma \b{r})$. Thus, $\Psi_\gamma$ belongs to the wave-function space ${\cal W}_{n_\gamma}^{\B_\gamma}$, and the scaling transformation in Eq.~(\ref{Psig}) defines a one-to-one map between ${\cal W}_n^\B$ and ${\cal W}_{n_\gamma}^{\B_\gamma}$. This leads to, for all $n \in {\cal D}^\B$,
\begin{eqnarray}
\gamma^2 F^{\l,\B}[n] &=& \gamma^2 \min_{\Psi \in {\cal W}^\B_n} \bra{\Psi} \hat{T} + \l \hat{W}_\text{ee} \ket{\Psi}
\nonumber\\
&=& \min_{\Psi \in {\cal W}^\B_n} \bra{\Psi_\gamma} \hat{T} + \l \gamma \hat{W}_\text{ee} \ket{\Psi_\gamma}
\nonumber\\
&=& \min_{\Psi \in {\cal W}^{\B_\gamma}_{n_\gamma}} \bra{\Psi} \hat{T} + \l \gamma \hat{W}_\text{ee} \ket{\Psi}
\nonumber\\
&=& F^{\l\gamma,\B_\gamma}[n_\gamma],
\label{}
\end{eqnarray}
and, for $\gamma=1/\l$,
\begin{eqnarray}
F^{\l,\B}[n] = \l^2 F^{\B_{1/\l}}[n_{1/\l}].
\label{}
\end{eqnarray}
This scaling relation generalizes for a finite basis set $\B$ the standard scaling relation $F^{\l}[n] = \l^2 F[n_{1/\l}]$. Hence, the scaling relation for the basis-set correction density functional $\bar{E}^{\l,\B}[n] = F^\l[n] - F^{\l,\B}[n]$ is
\begin{eqnarray}
\bar{E}^{\l,\B}[n] &=& \l^2 \bar{E}^{\B_{1/\l}}[n_{1/\l}].
\label{Elbscaling}
\end{eqnarray}
Equation~(\ref{Elbscaling}) can be used to introduce the dependence on $\l$ in any approximation that we have for $\bar{E}^{\B}$. Note that if we neglect the scaling of the density and of the basis set, we obtain 
\begin{eqnarray}
\bar{E}^{\l,\B}[n] &\approx& \l^2 \bar{E}^{\B}[n],
\label{Elbnoscaling}
\end{eqnarray}
which essentially corresponds to what was used by Mester and K\'allay~\cite{MesKal-JPCL-25}.

\subsubsection{Approximation for $\bar{E}^{\l,\B}$}
\label{sec:approxElB}

For the standard case $\l=1$, we will consider local approximations for the basis-set correction density functional $\bar{E}^{\B}$ of the form~\cite{LooPraSceTouGin-JPCL-19,HesGinReiKnoWerTou-JCC-24}
\begin{eqnarray}
\bar{E}^{\B}_\text{local}[n] &=& \int_{\mathbb{R}^3} \bar{e}^\sr_\text{c,md}(n(\b{r}),\nabla n(\b{r}), \mu^\B(\b{r})) \d \b{r},
\label{EBrholocal}
\end{eqnarray}
where $\bar{e}^\sr_\text{c,md}(n,\nabla n, \mu)$ is the so-called complementary multideterminant short-range correlation energy density. In Eq.~(\ref{EBrholocal}), $\mu^\B(\b{r})$ is a local range-separation parameter which provides a local measure of the incompleteness of the basis set $\B$. It is taken as~\cite{GinPraFerAssSavTou-JCP-18,LooPraSceTouGin-JPCL-19}
\begin{equation}
\mu^\B(\b{r}) = \frac{\sqrt{\pi}}{2} W^\B(\b{r}),
\end{equation}
where $W^\B(\b{r})$ is the on-top value of the effective interaction
\begin{equation}
W^\B(\b{r}) = 
\begin{cases}
\frac{f^\B(\b{r})}{n_{2}^\B(\b{r})}, & \text{if} \; n_{2}^\B(\b{r}) \neq 0,\\
\infty, & \text{otherwise},
\end{cases}
\label{WBr}
\end{equation}
with, assuming a basis set of real-valued orthonormal orbitals $\B = \{ \varphi_p \}_{p=1}^M$,
\begin{equation}
f^\B(\b{r}) = \sum_{pq}^{\text{all}} \sum_{rstu}^{\text{act}} w_{pqrs}^\B \Gamma_{rstu}^\B\varphi_p(\b{r}) \varphi_q(\b{r}) \varphi_t(\b{r}) \varphi_u(\b{r}),
\label{fBr}
\end{equation}
where $w_{pqrs}^\B=\braket{pq}{rs}$ are the two-electron Coulomb integrals, and
\begin{equation}
n_{2}^\B(\b{r}) = \sum_{pqrs}^{\text{act}} \Gamma_{pqrs}^\B\varphi_p(\b{r}) \varphi_q(\b{r}) \varphi_r(\b{r}) \varphi_s(\b{r}).
\label{n2Br}
\end{equation}
In Eqs.~(\ref{fBr}) and~(\ref{n2Br}), ``all'' means that the indices run over all (occupied + virtual) orbitals, ``act'' means that the indices run over all ``active'' orbitals (i.e., all occupied valence orbitals for the frozen-core version of the basis-set correction~\cite{GinSceLooTou-JCP-20}), and $\Gamma_{pqrs}^\B = 2 \bra{\Psi^\B_\text{loc}} \hat{a}_{r\downarrow}^\dagger \hat{a}_{s\uparrow}^\dagger \hat{a}_{q\uparrow} \hat{a}_{p\downarrow} \ket{\Psi^\B_\text{loc}}$ is the opposite-spin two-electron density matrix of a localizing wave function $\Psi^\B_\text{loc}$. The role of $\Psi^\B_\text{loc}$ is minor~\cite{GinPraFerAssSavTou-JCP-18,GinTraPraTou-JCP-21} so that in practice a single-determinant wave function in the basis set $\B$ can simply be used.

Upon scaling of the basis set, $\B_\gamma = \{ \varphi_{p,\gamma} \}_{p=1}^M$, the two-electron density matrix is invariant, $\Gamma_{pqrs}^{\B_\gamma}$ = $\Gamma_{pqrs}^\B$, and the two-electron integrals scales as $w_{pqrs}^{\B_\gamma} = \gamma w_{pqrs}^{\B}$. Thus, we have $f^{\B_\gamma}(\b{r}) = \gamma^7 f^\B(\gamma\b{r})$ and $n_{2}^{\B_\gamma}(\b{r}) = \gamma^6 n_{2}^\B(\gamma\b{r})$, and the scaling relation for the local range-separation parameter is 
\begin{equation}
\mu^{\B_\gamma}(\b{r}) = \gamma \mu^\B(\gamma\b{r}).
\end{equation}
Using Eq.~(\ref{Elbscaling}), we thus find the corresponding local approximation for $\bar{E}^{\l,\B}$
\begin{eqnarray}
\bar{E}^{\l,\B}_\text{local}[n] &=& \l^2 \int_{\mathbb{R}^3} \bar{e}^\sr_\text{c,md}(n_{1/\l}(\b{r}),\nabla_\b{r} n_{1/\l}(\b{r}), \mu^{\B_{1/\l}}(\b{r})) \d \b{r}
\nonumber\\
                   &=& \l^5 \int_{\mathbb{R}^3} \bar{e}^\sr_\text{c,md}\left(\frac{n(\b{r})}{\l^3},\frac{\nabla_\b{r} n(\b{r})}{\l^4}, \frac{\mu^\B(\b{r})}{\l} \right) \d \b{r},
\label{ElBlocalscaling}
\end{eqnarray}
where we have used the coordinate transformation $\b{r} \to \l \b{r}$.

\begingroup
\squeezetable
\begin{table*}[t]
\caption{MAEs and MEs (in kcal/mol) with respect to the CBS limit on the AE6 test set with vdz, vtz, vqz, and v5z basis sets for the DS1DH-PBE and 1DH-BLYP double-hybrid approximations, without and with the basis-set correction (with and without density scaling, i.e. DSBSC and BSC), and without and with the basis-set one-electron correction (OEC). All the results are for a value of $\lambda =0.65$.}
\label{tab:AE6}
\begin{tabular}{lccccccccc}
\hline
\hline
          & \multicolumn{9}{c}{AE6}                                                       \\
          \cline{2-10}
          & \multicolumn{4}{c}{MAE}           &                 &  \multicolumn{4}{c}{ME}  \\
          \cline{2-5} \cline{7-10}
Method    &vdz  &vtz  &vqz  &v5z              & \phantom{xxxx}  & vdz & vtz & vqz & v5z  \\
\hline
DS1DH-PBE           &23.42 &6.70 &2.57 &1.25 & &-23.42  &-6.70  &-2.57 &-1.25 \\
DS1DH-PBE+DSBSC     &5.76 &0.70 &0.77 &0.61 & &-5.76 &0.05 &0.77 &0.61 \\
DS1DH-PBE+DSBSC+OEC &3.52 &1.14 &0.89 &0.61 & &3.38 &1.02 &0.89 &0.61 \\
DS1DH-PBE+BSC       &9.02 &1.30 &0.26 &0.25 & &-9.02 &-1.30 &0.09 &0.22 \\
DS1DH-PBE+BSC+OEC   &1.27 &0.49 &0.32 &0.25 & &0.12 &-0.32 &0.21 &0.22 \\[0.1cm]
1DH-BLYP            &24.06 &6.74 &2.53 &1.25 & &-24.06 &-6.74 &-2.53 &-1.25 \\
1DH-BLYP+DSBSC      &6.41 &0.61 &0.81 &0.61 & &-6.41 &0.00 &0.81 &0.61 \\
1DH-BLYP+DSBSC+OEC  &3.47 &1.15 &0.90 &0.61 & &3.34 &1.04 &0.89 &0.61 \\
1DH-BLYP+BSC        &9.67 &1.34 &0.24 &0.25 & &-9.67 &-1.34 &0.13 &0.22 \\
1DH-BLYP+BSC+OEC    &1.31 &0.47 &0.32 &0.25 & &0.08 &-0.30 & 0.21 &0.22 \\
\hline
\hline
\end{tabular}
\end{table*}
\endgroup

\begingroup
\squeezetable
\begin{table*}[t]
\caption{MAEs and MEs (in kcal/mol) with respect to the CBS limit on the BH6 test set with vdz, vtz, vqz, and v5z basis sets for the DS1DH-PBE and 1DH-BLYP double-hybrid approximations, without and with the basis-set correction (with and without density scaling, i.e. DSBSC and BSC), and without and with the basis-set one-electron correction (OEC). All the results are for a value of $\lambda =0.65$.}
\label{tab:BH6}
\begin{tabular}{lccccccccc}
\hline
\hline
          & \multicolumn{9}{c}{BH6}                                                       \\
          \cline{2-10}
          & \multicolumn{4}{c}{MAE}           &                 &  \multicolumn{4}{c}{ME}  \\
          \cline{2-5} \cline{7-10}
Method    &vdz  &vtz  &vqz  &v5z              & \phantom{xxxx}  & vdz & vtz & vqz & v5z  \\
\hline
DS1DH-PBE           &1.97 &0.83 &0.32 &0.12 & &-0.11 &0.17 &0.08 &0.09 \\
DS1DH-PBE+DSBSC     &1.49 &0.68 &0.33 &0.16 & &-1.06 &-0.22 &-0.13 &-0.04 \\
DS1DH-PBE+DSBSC+OEC &0.93 &0.45 &0.26 &0.16 & &0.33 &0.05 &-0.04 &-0.04 \\
DS1DH-PBE+BSC       &1.40 &0.61 &0.29 &0.13 & &-0.93 &-0.17 &-0.10 &-0.02 \\
DS1DH-PBE+BSC+OEC   &0.88 &0.38 &0.21 &0.13 & &0.46 &0.10 &-0.01 &-0.02 \\[0.1cm]
1DH-BLYP            &1.98 &0.85 &0.34 &0.13 & &-0.30 &0.13 &0.07 &0.09 \\
1DH-BLYP+DSBSC      &1.55 &0.70 &0.34 &0.15 & &-1.27 &-0.27 &-0.15 &-0.04 \\
1DH-BLYP+DSBSC+OEC  &0.92 &0.44 &0.25 &0.15 & &0.31 &0.04 &-0.04 &-0.04 \\
1DH-BLYP+BSC        &1.43 &0.64 &0.30 &0.12 & &-1.14 &-0.22 &-0.12 &-0.03 \\
1DH-BLYP+BSC+OEC    &0.87 &0.38 &0.21 &0.12 & &0.44 &0.09 &-0.02 &-0.03  \\
\hline
\hline
\end{tabular}
\end{table*}
\endgroup

\section{Computational details}

We have extended the implementation of the DBBSC method reported in Ref.~\onlinecite{HesGinReiKnoWerTou-JCC-24} in the software MOLPRO~\cite{WerKnoKniManSch-WIR-12,WerKnoManBlaDolHesKatKohKorKreMaMilMitPetPolRauSib-JCP-20,Molpro-PROG-25} to include the density-scaled basis-set correction (DSBSC) in Eq.~(\ref{ElBlocalscaling}) for the one-parameter double-hybrid approximations~\cite{ShaTouSav-JCP-11}. We consider two such one-parameter double-hybrid approximations: DS1DH-PBE~\cite{ShaTouSav-JCP-11}, based on the Perdew-Burke-Ernzerhof (PBE)~\cite{PerBurErn-PRL-96} exchange and correlation density functionals and including density scaling in the correlation functional, and 1DH-BLYP~\cite{ShaTouSav-JCP-11}, based on the Becke~\cite{Bec-PRA-88} exchange density functional and the Lee-Yang-Parr (LYP)~\cite{LeeYanPar-PRB-88} correlation density functional without including density scaling in the correlation functional. In both cases, we use a parameter of $\lambda = 0.65$, as recommended in Ref.~\onlinecite{ShaTouSav-JCP-11}. For the basis-set correction functional, we use the approximation based on the PBE correlation functional given in Ref.~\onlinecite{LooPraSceTouGin-JPCL-19}. Instead of including the basis-set correction self-consistently, we first perform a standard DS1DH-PBE or 1DH-BLYP calculation, and then add a posteriori the DSBSC functional evaluated with the DS1H-PBE or 1H-BLYP single-determinant wave function (for both the density and the localizing wave function in the calculation of the local range-separation parameter, see Section~\ref{sec:approxElB}). We refer to these calculations as DS1DH-PBE+DSBSC or 1DH-BLYP+DSBSC. We also test neglecting the density scaling in the basis-set correction (BSC) functional [Eq.~(\ref{Elbnoscaling})]. We refer to the corresponding calculations as DS1DH-PBE+BSC or 1DH-BLYP+BSC. 

We use the Dunning correlation-consistent polarized valence basis sets~\cite{Dun-JCP-89} cc-pVDZ,  cc-pVTZ, cc-pVQZ, and cc-pV5Z (abbreviated as vdz, vtz, vqz, and v5z). The MP2 contributions in the double-hybrid calculations are done with the frozen-core approximation. Accordingly, we use the frozen-core version~\cite{GinSceLooTou-JCP-20,HesGinReiKnoWerTou-JCC-24} of the basis-set correction, which means that the active orbitals in Eqs.~(\ref{fBr}) and Eqs.~(\ref{n2Br}) are limited to the occupied valence orbitals and the basis-set correction functional is evaluated with the valence-only density. Since our basis-set correction functional only takes into account the two-electron basis-set incompleteness error, we also test the addition of a one-electron basis-set correction which simply consists in the energy difference between the single-determinant DS1H-PBE or 1H-BLYP calculation in the largest basis set v5z and the corresponding calculation in the considered basis set $\B$ (similarly to what was done at the HF level in Refs.~\onlinecite{LooPraSceTouGin-JPCL-19,TraAdjFenLygMadPosHamPerTouGinPiq-CC-24}). We refer to the corresponding calculations including this one-electron correction (OEC) as DS1DH-PBE+DSBSC+OEC or 1DH-BLYP+DSBSC+OEC when using density scaling in the basis-set correction functional, and DS1DH-PBE+BSC+OEC or 1DH-BLYP+BSC+OEC when neglecting density scaling in the basis-set correction functional. This one-electron basis-set correction should give very similar results as the complementary auxiliary basis-set (CABS) single-excitation correction~\cite{MesKal-JCTC-23,HesGinReiKnoWerTou-JCC-24,MesNagKal-JCTC-24,MesKal-CPL-25,MesKal-JPCL-25} and is technically simpler to apply for the present calculations.

We perform calculations on the AE6 and BH6 datasets~\cite{LynTru-JPCA-03}. The AE6 dataset is a small representative benchmark of six atomization energies consisting of SiH$_4$, S$_2$, SiO, C$_3$H$_4$ (propyne), C$_2$H$_2$O$_2$ (glyoxal), and C$_4$H$_8$ (cyclobutane). The BH6 dataset is a small representative benchmark of forward and reverse hydrogen transfer barrier heights of three reactions, OH + CH$_4$ $\rightarrow$ CH$_3$ + H$_2$O, H + OH $\rightarrow$ O + H$_2$, and H + H$_2$S $\rightarrow$ HS + H$_2$. All the calculations for the AE6 and BH6 datasets are performed at the geometries optimized by quadratic configuration interaction with single and double excitations with the modified Gaussian-3 basis set (QCISD/MG3)~\cite{Minnesota-DATABASE-XX}. We estimate the CBS limit of the double-hybrid approximations by standard basis-set extrapolation: we use the single-determinant results with the largest basis set v5z~\cite{HalHelJorKloOls-CPL-99} and we use a two-point $X^{-3}$ extrapolation of the MP2 correlation energy with the vqz ($X=4$) and v5z ($X=5$) basis sets~\cite{HelKloKocNog-JCP-97,HalHelJorKloKocOlsWil-CPL-98}. We compute mean absolute errors (MAEs) and mean errors (MEs) with respect to the CBS-limit estimates.

\begin{figure*}
\includegraphics[scale=0.30,angle=-90]{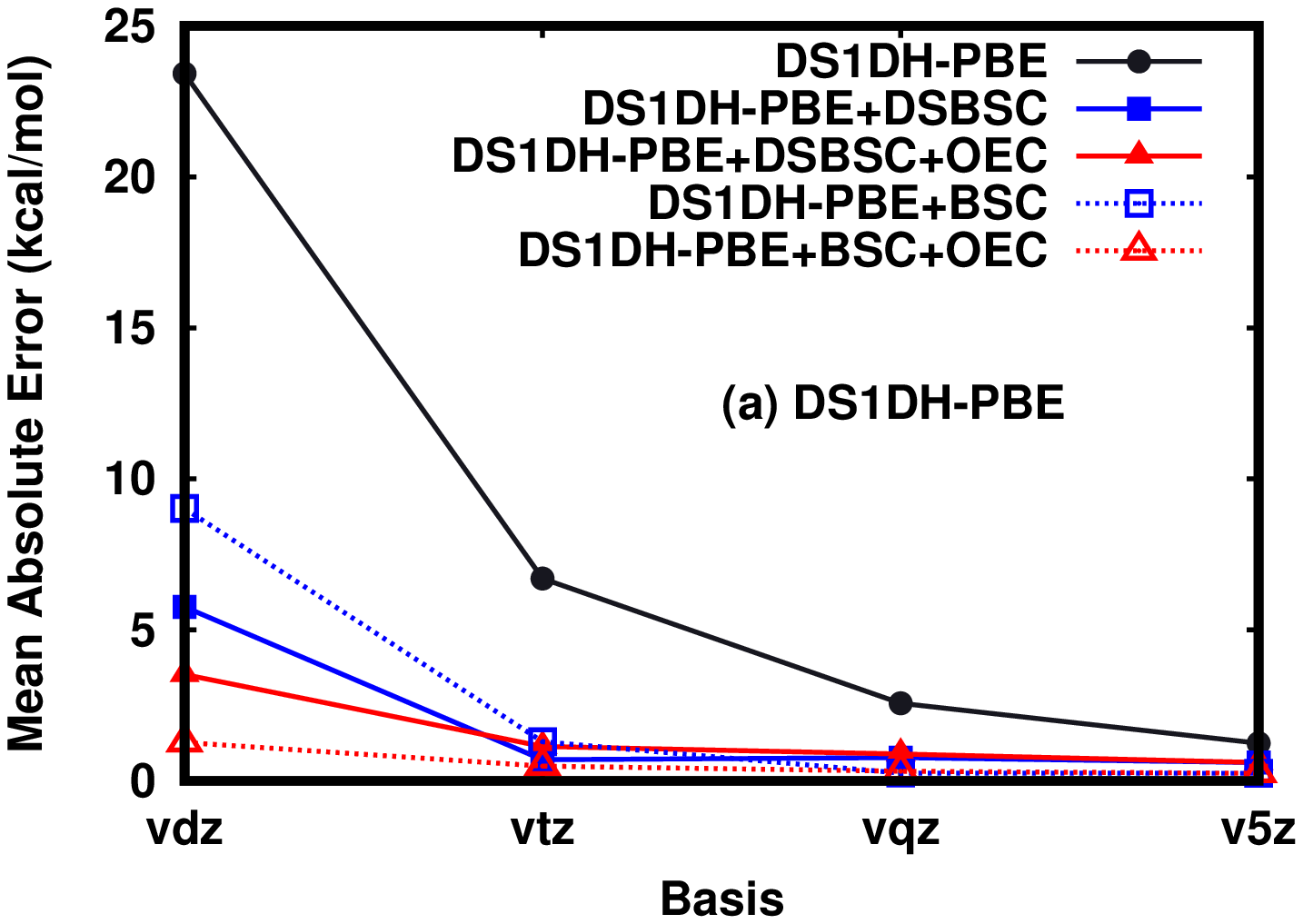}
\includegraphics[scale=0.30,angle=-90]{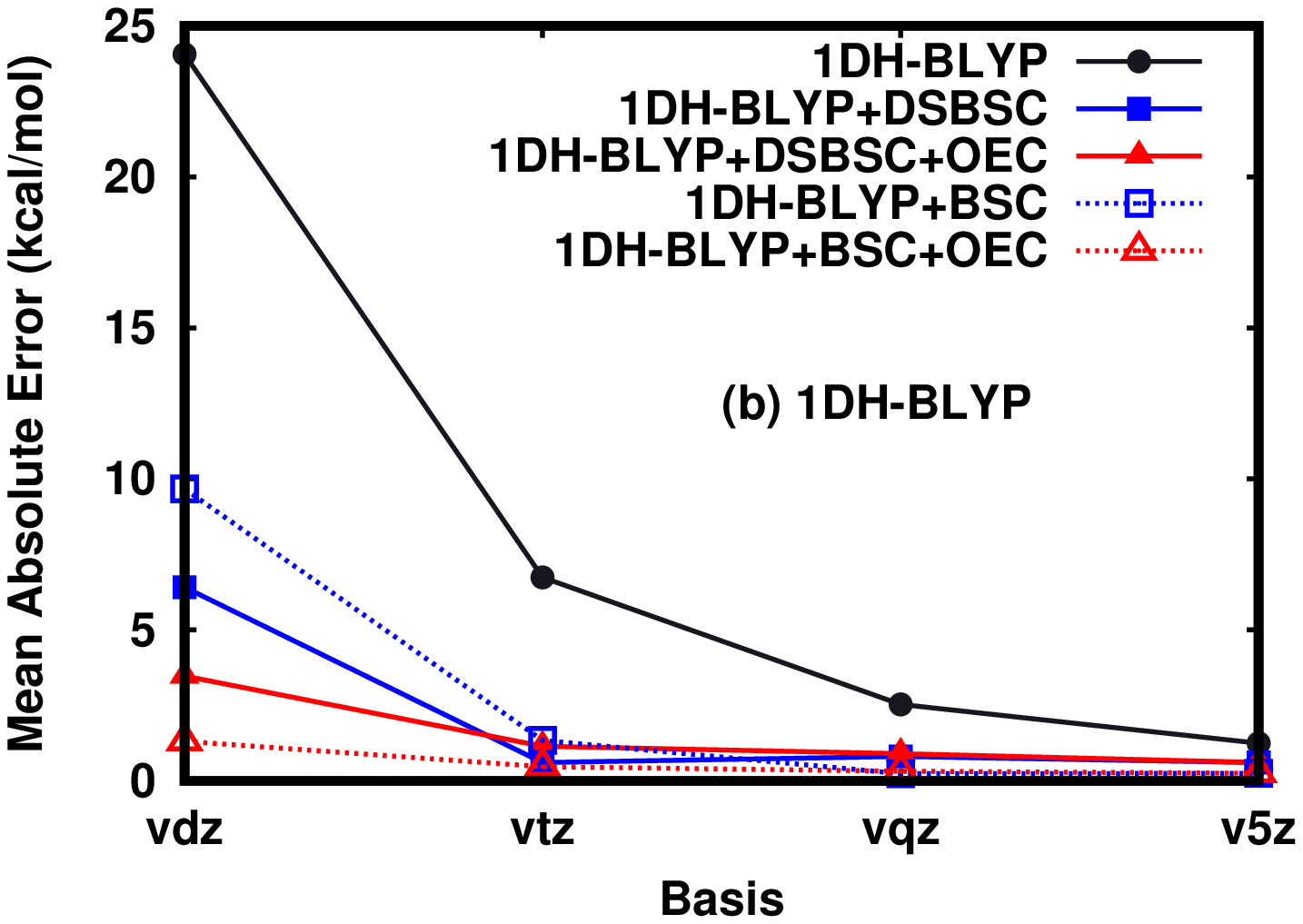}
\caption{MAEs (in kcal/mol) with respect to the CBS limit on the AE6 test set as functions of basis sets for the (a) DS1DH-PBE and (b) 1DH-BLYP double-hybrid approximations, without and with the basis-set correction (with and without density scaling, i.e. DSBSC and BSC), and without and with the basis-set one-electron correction (OEC). All the results are for a value of $\lambda =0.65$.}
\label{fig:AE6_MAE_PBE_BLYP}
\end{figure*}

\begin{figure*}
\includegraphics[scale=0.30,angle=-90]{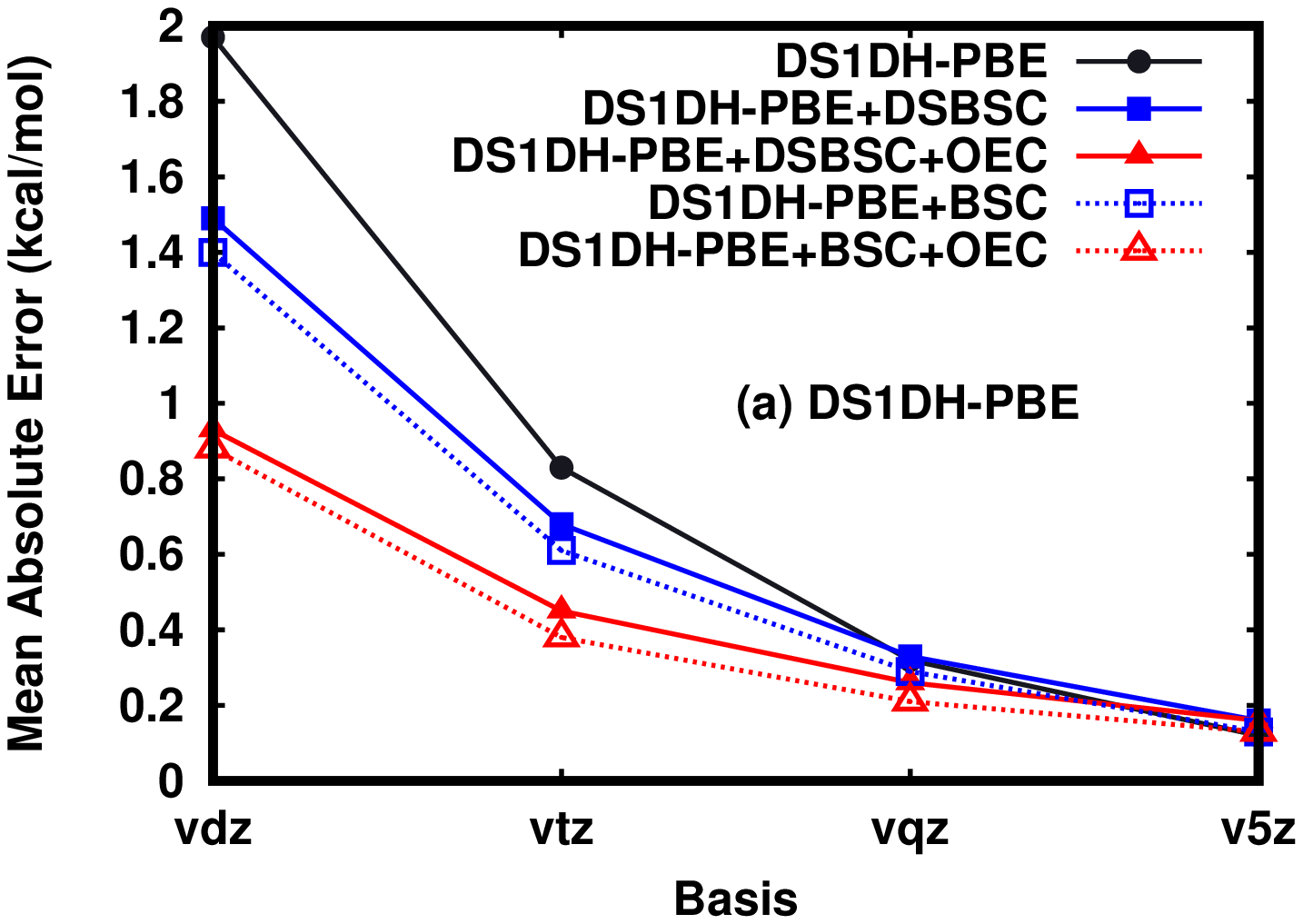}
\includegraphics[scale=0.30,angle=-90]{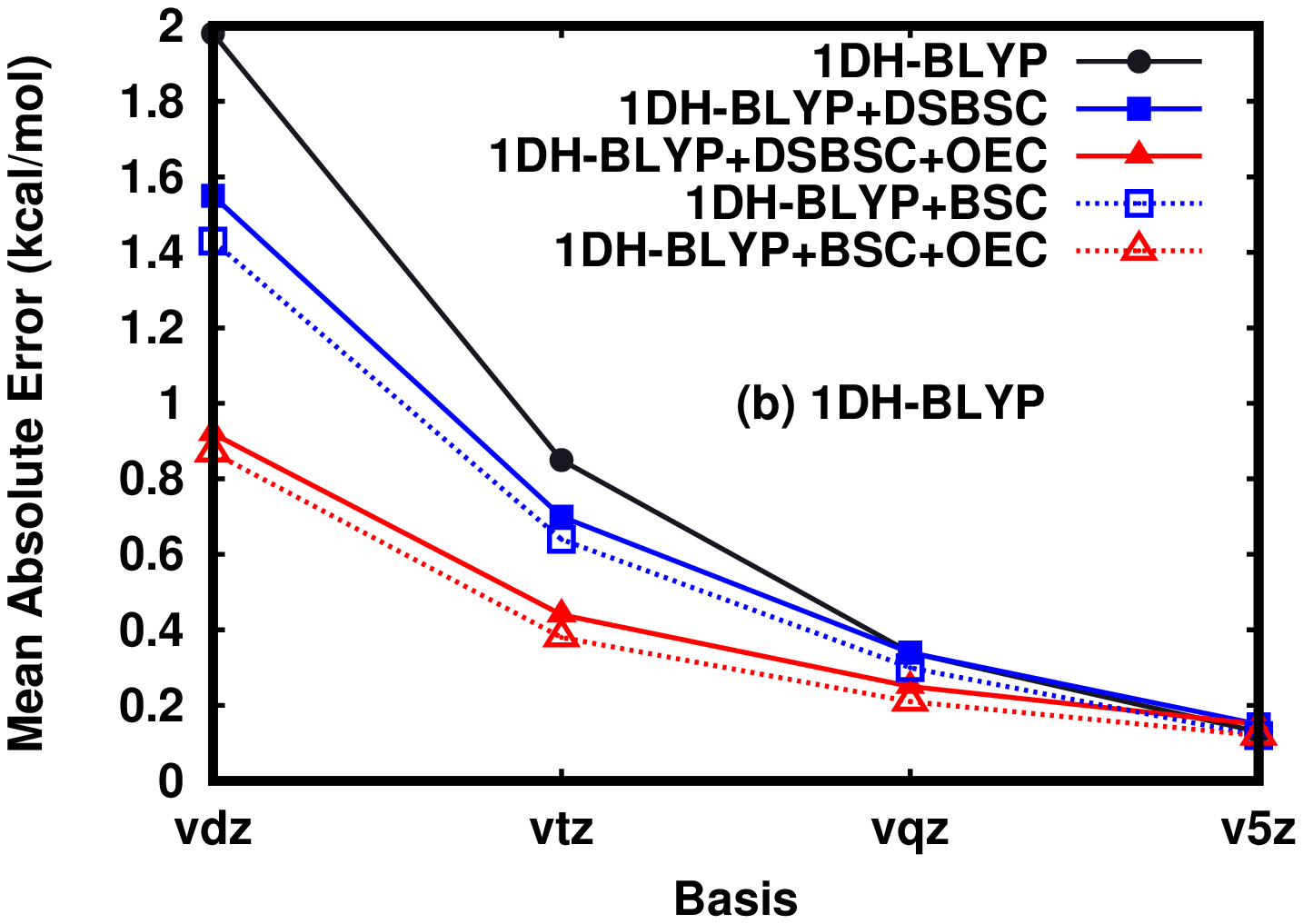}
\caption{MAEs (in kcal/mol) with respect to the CBS limit on the BH6 test set as functions of basis sets for the (a) DS1DH-PBE and (b) 1DH-BLYP double-hybrid approximations, without and with the basis-set correction (with and without density scaling, i.e. DSBSC and BSC), and without and with the basis-set one-electron correction (OEC). All the results are for a value of $\lambda =0.65$.}
\label{fig:BH6_MAE_PBE_BLYP}
\end{figure*}

\section{Results and discussion}

The MAEs and MEs with respect to the CBS limit on the AE6 and BH6 test sets for the different methods are reported in Table~\ref{tab:AE6} and~\ref{tab:BH6}. The MAEs are also plotted as functions of the basis sets in Figure~\ref{fig:AE6_MAE_PBE_BLYP} and~\ref{fig:BH6_MAE_PBE_BLYP}. 

We start by discussing the results for the AE6 test set. The atomization energies calculated with the DS1DH-PBE double-hybrid approximation have a fairly large dependence on the size of the basis set. The atomization energies are systematically underestimated with respect to the CBS limit, with a ME of -23.42 kcal/mol for the vdz basis set and a ME of 1.25 kcal/mol for the v5z basis set. Adding the basis-set correction functional with density scaling greatly accelerates the basis convergence: DS1DH-PBE+DSBSC gives a MAE of 5.76 kcal/mol for the vdz basis set, and a MAE below 0.8 kcal/mol already with the vtz basis set. While DS1DH-PBE+DSBSC still gives underestimated atomization energies with respect to the CBS limit for the vdz basis set, it gives overestimated atomization energies with respect to the CBS limit starting from the vtz basis set. This means that DSBSC tends to overcorrect a bit the basis-set error for the vtz basis set and larger basis sets. Adding the one-electron basis-set correction further helps to reduce the basis-set error by about 2 kcal/mol for the vdz basis set, although it tends to further overcorrect a bit for the vtz and vqz basis sets. Neglecting the density scaling in the basis-set correction functional tends to reduce the magnitude of the correction. The consequence is that DS1DH-PBE+BSC gives larger MAEs for the vdz and vtz basis sets (9.02 and 1.30 kcal/mol, respectively) but smaller MAEs for the vqz and v5z basis sets (0.26 and 0.25 kcal/mol, respectively) since the overcorrection of DSBSC for large basis sets is reduced. Combining the basis-set correction functional without density scaling and the one-electron basis-set correction appears to lead to an optimal compensation of errors: DS1DH-PBE+BSC+OEC gives a MAE of 1.27 kcal/mol for the vdz basis set, and MAEs below 0.5 kcal/mol for larger basis sets. Starting from the 1DH-BLYP double-hybrid approximation gives very similar results, resulting in identical conclusions.

We discuss now the results for the BH6 test set. The reaction barrier heights calculated with the double-hybrid approximations have much smaller basis-set errors. DS1DH-PBE gives a MAE of 1.97 kcal/mol for the vdz basis set, and MAEs below 1 kcal/mol for larger basis sets. Overall, adding the basis-set correction functional with density scaling accelerates the basis convergence, but to a lesser extend compared to what was observed on the AE6 test set. For the v5z basis set, DS1DH-PBE+DSBSC gives in fact a slightly larger MAE (0.16 kcal/mol) than DS1DH-PBE (0.12 kcal/mol). This is due to the fact that, again, DSBSC tends to overcorrect a bit the basis-set error for the large basis sets. Adding the one-electron basis-set correction significantly reduces the basis-set error for the vdz, vtz, and vqz basis sets. On this test set, neglecting the density scaling in the basis-set correction functional always leads to slightly smaller MAEs, whether or not the one-electron basis-set correction is present: DS1DH-PBE+BSC and DS1DH-PBE+BSC+OEC have on average slightly smaller basis-set errors than DS1DH-PBE+DSBSC and DS1DH-PBE+DSBSC+OEC, respectively. In particular, neglecting the density scaling in the basis-set correction functional reduces the overcorrection of DSBSC for the v5z basis set. Like for the AE6 test set, it appears that combining the basis-set correction functional without density scaling and the one-electron basis-set correction leads to the smallest MAEs:  DS1DH-PBE+BSC+OEC gives a MAE of 0.88 kcal/mol for the vdz basis set, and MAEs below 0.4 kcal/mol for larger basis sets. Here as well, starting from the 1DH-BLYP double-hybrid approximation gives very similar results.

\section{Conclusion}

In this work, we have developed the theory justifying the application of the DBBSC approach to double-hybrid approximations. More specifically, we showed how to extend the basis-set correction functional to the context of the one-parameter double-hybrid DFT framework of Ref.~\onlinecite{ShaTouSav-JCP-11}. The exact dependence of the basis-set correction functional on the coupling-constant parameter $\l$ involves a uniform coordinate scaling by a factor $1/\l$ of the density and of the basis functions. Neglecting this uniform coordinate scaling corresponds essentially to the recent work of Mester and K\'allay~\cite{MesKal-JPCL-25} on the application of the DBBSC method to double-hybrid approximations. Thus, the present work provides a clear theoretical framework to understand the work of Ref.~\onlinecite{MesKal-JPCL-25}.

Calculations on the AE6 and BH6 datasets showed that the DBBSC method efficiently accelerates the basis convergence of one-parameter double-hybrids. Unexpectedly, we found that neglecting the uniform coordinate scaling in the basis-set correction functional and simultaneously adding a one-electron basis-set correction leads to an optimal compensation of errors, with atomization energies and reaction barrier heights within roughly chemical accuracy (1 kcal/mol) on average of the double-hybrid CBS-limit values with a basis set as small as vdz. This validates the strategy of Mester and K\'allay~\cite{MesKal-JPCL-25} and confirms their findings.

Even though neglecting the uniform coordinate scaling in the basis-set correction functional appears to be adequate for MP2-based double hybrids and for the local PBE-based basis-set correction functional for the calculations of atomization energies and reaction barrier heights, it could be worthwhile to recheck the effect of the uniform coordinate scaling when other types of double hybrids are used (such as the ones based on the random-phase approximation), when other types of approximate basis-set correction functionals are used (such as the ones based on a different correlation functional~\cite{HesGinReiKnoWerTou-JCC-24} or the ones involving the on-top pair density~\cite{GinSceTouLoo-JCP-19,GinSceLooTou-JCP-20}), or when calculating other properties.


\end{document}